\documentclass[preprint,12pt]{elsarticle}

\usepackage{lineno,hyperref}
\usepackage{amsmath}
\usepackage{hyperref}
\usepackage[section]{placeins}
\modulolinenumbers[5]

\journal{Nucl. Instr. Meth. Phys. Res. - Section B}

\begin{document}

\begin{frontmatter}

\title{Self-consistent ion beam analysis: \\ an approach by multi-objective optimization}

\author{T. F. Silva\fnref{myfootnote}}
\fntext[myfootnote]{Corresponding author. e-mail: tfsilva@if.usp.br}
\author{C.L. Rodrigues}
\author{N. Added}
\author{M.A. Rizzutto}
\author{M.H. Tabacniks}
\address{Instituto de Física da Universidade de S\~ao Paulo, Rua do matão, trav. R 187,\\ 05508-090 S\~ao Paulo, Brazil.}
\author{T. H\"oschen}
\author{U. von Toussaint}
\author{M. Mayer}
\address{Max-Planck-Institut f\"ur Plasmaphysik, Boltzmannstr. 2, D-85748 Garching, Germany.}

\begin{abstract}
  Ion Beam Analysis (IBA) comprises a set of analytical techniques suited for material analysis, many of which are rather closely related. Self-consistent analysis of several IBA techniques takes advantage of this close relationship to combine different Ion Beam measurements in a unique model to obtain an improved characterization of the sample. This approach provides a powerful tool to obtain an unequivocal and reliable model of the sample, increasing confidence and reducing ambiguities. Taking advantage of the recognized reliability and quality of the simulations provided by SIMNRA, we developed a multi-process program for a self-consistent analysis based on SIMNRA calculations. MultiSIMNRA uses computational algorithms to minimize an objective function running multiple instances of SIMNRA. With four different optimization algorithms, the code can handle sample and setup parameters (including correlations and constraints), to find the set of parameters that best fits simultaneously all experimental data.
\end{abstract}


\end{frontmatter}



\section{Introduction}

Ion Beam Analysis (IBA) comprises a set of well-established analytical techniques for material characterization where the interest lies on the surface or near-surface region \cite{JEYNES_centuryofiba}. These techniques are valuable for many fields of material science, finding applications in semiconductors, sensors, magnetics, coatings, and also to other areas such as biology, environmental monitoring and cultural heritage.

Computational tools played an important role in the data interpretation during the historical development of IBA techniques \cite{RAUHALA_statusofsoftware}. They are essential for those techniques based on particle scattering spectrometry, such as Rutherford Backscattering Spectrometry (RBS), Elastic Backscattering Spectrometry (EBS), Elastic Recoil Detection Analysis (ERDA) and Nuclear Reaction Analysis (NRA). The close relationship between computational tools and scattering techniques began in the early 1970s when Ziegler and Baglin reported, for the first time, the use of computational modeling of many physical phenomena for scattering analysis \cite{RAUHALA_statusofsoftware,ZIGLER_seminal}. Since then, many advances occurred in the comprehension of physical processes and measurement system effects, improving computational modeling \cite{MAYER_computersimulations}. 

A well succeeded case is the SIMNRA software \cite{MAYER_simnra}, which is a computer program widely adopted by the IBA community. Its success is due to its detailed physical modeling coupled to an intuitive user interface, simplified use, optimization features and the simple interface that enables the use of non-Rutherford cross-sections provided by IBANDL database \cite{ABRIOLA_referencedatabase} and SigmaCalc \cite{GURBICH_evaluatedcrosssection}. The strength of SIMNRA lies on trusted modeling of the physical processes involved in the scattering calculation and measurements system effects. More recent upgrades concerned the skewness of energy spread distributions, improved handling of reaction cross-sections with fine structure, generalized layer roughness, and sample porosity \cite{MAYER_simnra7}.

Other software packages have been developed including algorithms capable of treating complex experimental conditions, some of which also include advanced levels of automation \cite{RAUHALA_statusofsoftware}. An important example is DataFurnance \cite{BARRADAS_advancedphysics}, which uses advanced algorithms to minimize an objective function to fit the sample model to the experimental data. The DataFurnance software is able to handle simultaneously and self-consistently different IBA spectra obtained by multiple techniques applied in the same sample \cite{BARRADAS_simultaneousandconsistent}. 

Butler has shown that a single RBS spectrum may lead to degenerated solutions in some cases \cite{BUTLER_criteria} and Alkemade argues that, to avoid all ambiguities, it may be necessary to measure up to $(N - 1)$ independent spectra to characterize N parameters of a sample \cite{ALKEMADE_ambiguity}. The self-consistent treatment of the data ensures that the information contained in all spectra contribute to the sample characterization, taking advantage of the complementarity of different IBA techniques \cite{JEYNES_totaliba}. This approach enables the reliable and unequivocal modeling of a sample, but equally important, the self-consistent analysis inherits the accuracy of the most accurate component of the analysis \cite{JEYNES_accuratedetermination, COLAUX_highaccuracy}.  In that sense, high accuracy together with self-consistent analysis seems to be the road to the future of IBA to stay competitive with other techniques.

Taking advantage of the recognized reliability and quality of the simulations provided by SIMNRA, we developed a multi-process program for a self-consistent analysis based on SIMNRA calculations. MultiSIMNRA uses computational algorithms to minimize an objective function running at the same time multiple instances of SIMNRA. With four different optimization algorithms available, the code can handle sample and setup parameters (including correlations and constraints), to find the set of parameters that best fits simultaneously all experimental data. Completing five years since the release of the first version\cite{Multisimnra}, many lessons have been learned with use, collaborations and feedback from users worldwide. This publication summarizes the current status and implemented advances in the data process and handling. MultiSIMNRA has been tested in several situations \cite{app_1,app_2,app_3,app_4,app_5,app_6,app_7,app_8,app_9,app_10,app_11}, always providing quality information constrained by multiple ion beam techniques. To illustrate how powerful this approach can be, we present results for some selected cases and discuss some interesting features of our code. We show how MultiSIMNRA helps automating simple tasks and how a difficult analysis can be simplified.

\section{Methods}

In a self-consistent approach, different IBA techniques are used to characterize the elements in a sample. The approach is based on the combination of several ion beam measurements in a unique model, where experimental data of all measurements are fitted simultaneously for the same sample description. The simultaneous fit of all the spectra implies that information obtained from each technique is used as a boundary condition, wherever applicable on the others. In this way, all information about the sample emerges consistently from the set of spectra analyzed.

The MultiSIMNRA software uses multiple instances of the SIMNRA to calculate simulated spectra for different experimental conditions, and compares with experimental data using an objective function (figure of merit function). The fitting procedure is made by the minimization of an objective function modifying the fitting parameters accordingly to an optimization algorithm. It is considered as final result the model that presents the best possible match to the experimental data (i.e., the model that minimizes the global objective function). At the end of the optimization process an uncertainty estimation process is started. This procedure may be considered as a generalization of the reverse Monte-Carlo method \cite{mcgreevy_reverse_1988} or the backward induction method.

\subsection{Objective function}

MultiSIMNRA uses an objective function based on the calculation of quadratic differences ($\chi^2$ function). However, the simple combination of the standard $\chi^2$ function for each spectrum can result in a biased analysis if there are unaccounted model deviations since spectra with higher counting statistics receive a higher weight in the minimization process \cite{marler_survey_2004}. In addition, in numerical experiments it has also been observed that an adjusted weighting of the individual spectra results in faster convergence. Differences in counting statistics can occur either because of different experimental conditions (for example different solid angles for different detectors or different accumulated charges in different measurements) or because of different cross-sections involved in the physical processes \cite{marler_weighted_2010, deb_introducing_2006}. We compensate this unbalance by normalizing the  $\chi^2$ function by its expected value for each spectrum.

\begin{equation}
\left\langle \sum_{channels} \frac{\left( T_i - M_i \right)^2}{\sigma_i^2} \right\rangle = \left\langle \sum_{channels} \frac{\left( T_i - M_i \right)^2}{max(M_i,2)} \right\rangle = DoF
\end{equation}
where $max(M_i,2)$ is used to avoid the null uncertainties.

The $\chi^2$ definition for minimization using MultiSIMNRA is:

\begin{equation}
\label{MS_OF}
\chi^2_{red} = \frac{1}{S}\sum _{s=1}^{S} \left (   { \frac{a_s}{N_s-P_s-P_p} \cdot \sum_{channels} {\frac{(T_i-M_i)^2}{max(M_i,2)}}} \right ) 
\end{equation}
where $M_i$ and $T_i$ are respectively the calculated and experimental value of counts in channel $i$ of a spectrum $s$, $S$ is the total number of spectra, $N_s$ is the number of channels in the fitting region of the spectrum $s$, $P_s$ is the number of fitting parameters in the setup configuration for the measurement of the spectrum $s$, and $P_p$ is the number of fitting parameter in the sample depth profile (note that $DoF = N_s-P_s-P_p$ for each spectrum). MultiSIMNRA also enables the user to change the parameter $a_s$ (which is equal to one by default) which is a parameter that serves to provide a intentional unbalancing to the contribution of a certain spectrum to the total $\chi^2$.

Our choice of an objective function based on reduced $\chi^2$ instead of the $\chi^2$ itself can be justified since each spectrum has a different expected value for the latter. Thus, each spectrum contributes differently to the global value of an objective function based on the $\chi^2$ calculation. This difference affects the balance of the objective function attributing different statistical weights to each spectra \cite{marler_weighted_2010}. However, an objective function based on the reduced form of the $\chi^2$, e.g. normalized by the number of degrees-of-freedom, ensures that all the spectra have the same expected value (equals to  unity) and consequently the same contribution to the global objective function. Thus, the use of the mean reduced $\chi^2$ as the objective function ensures that all spectra in the analysis have the same statistical weights in the minimization process, and that the expected value of Eq. \ref{MS_OF} is also equal to one, independently on the number of spectra in the joint analysis. 

Formally, this is a multi-objective optimization problem solved by the weighted-sum method \cite{marler_survey_2004}. By adopting the weights to normalize the individual objective functions we equalize the scale of comparison provided by the global objective function. On this terms, the $a_s$ of eq. \ref{MS_OF} can be interpreted as a way to express preferences in the analysis, whilst leaving an arbitrary weight ($a_s$) to the analyst to be used as an 'ad-hoc' quality parameter. A higher $a_s$ represents a higher preference \cite{marler_weighted_2010}.

This problem was addressed differently by Barradas et. al\cite{BARRADAS_unambiguous}, who introduced a weight for each partial $\chi^2$ in the total $\chi^2$ calculation, in an attempt to correct for this bias. The basic idea is that, by assuming an objective function proportional to the squared difference of the experimental data and the simulated model, it enables the calculation of its expected value for each spectrum. This can be done as:

\begin{equation}
\left\langle \sum_{channels} \left( T_i - M_i \right)^2 \right\rangle = \left\langle \sum_{channels} \sigma_i^2 \right\rangle = \sum_{channels} M_i = I_{exp}
\end{equation}

The authors argued that in a less-than-perfect fit, the quadratic difference grows faster than proportional to $I_{exp}$, so they introduced a weight for each spectrum of $I_{exp}^{1.5}$ in an attempt to balance the objective function. It is worth to point out that, by doing so, the authors introduced a biased term that privileges spectra with a lower counting statistics, since the punishment in the objective function is stronger for spectra with higher integrated counts than for spectra with lower integrated counts (when compared to the unbiased weight of $I_{exp}$). The authors also included an extra term as a penalty for the increment of the number of layers ($n_{lay}$) in the sample description \cite{BARRADAS_datafurnance}. In this case, the combination of the partial $\chi^2$ for a balanced objective function is as follows:


\begin{equation}
\label{NDF_OF}
\chi^2_{NDF} = \sum _{s=1}^{S} \left (   \frac{\sum_{channels} {(M_i-T_i)^2}}{I_{exp}^{1.5} \cdot ( 1 + 0.01 \cdot n_{lay})}   \right ) 
\end{equation}

The NDF's objective function is also implemented in MultiSIMNRA for comparison reasons.

Other objective functions may be available in NDF, mainly for the Bayesian inference method of uncertainty estimation \cite{Nuno}. We refer to eq. \ref{NDF_OF} as an alternative example, and as the only version published until now for the NDF's objective function.

One should keep in mind that due to noise, imperfections in physical modeling, measuring artifacts, etc. it may not exist a unique set of parameters that minimizes the global and all the individual objective functions simultaneously. The Pareto optimal condition defines there is a set of parameters that minimize the global objective function, even if not presenting an optimized solution for at least one of the single objective functions. Therefore, the Pareto optimal is the best solution possible within the average agreement of all measurements, and the choice of the weights may affect the Pareto optimal \cite{marler_survey_2004}. 

Perhaps, this is the most critical argument in favor of adopting the weighted-sum method for this multi-objective optimization. Suppose all different spectra have about the same weight when combining the different measurements. In that case, the unknown or unaccounted systematic errors are averaged in the analysis, and no particular deviation of the forward model is preferred over the other. Otherwise, using an unbalanced objective function, spectra with higher statistics may dominate the analysis, even though they may have more flawed theoretical understanding, while neglecting those with fewer counts but possibly better modeled.

\subsection{Uncertainty calculation}

At the end of the minimization process, our code can also perform a statistically sound evaluation of the uncertainty of the fitted parameters. This evaluation is done using a Monte-Carlo calculation to explore the values of  $\chi^2$ in the neighborhood of all fitted parameters. For the determination of a confidence region with $1-\gamma$ confidence level, random variations of the parameters are accepted only if the changes in the $\chi^2$ are lower than a critical value ($\chi^2_{\gamma}$) determined by the $\chi^2$ probability density function and the confidence level ($1-\gamma$)~\cite{Cowan}. The accepted values are stored in a buffer file to account for the statistics in the near region of the Pareto optimal. The confidence level is determined by the user, and considered as 68.3\% (one standard deviation) by default.

The critical value $\chi^2_{\gamma}$ is defined by the equation:

\begin{equation*}
\int_0^{\chi_{\gamma}} f(z;n) dz = 1-\gamma
\end{equation*}
where $f(z;n)$ is the $\chi^2$ probability density function, $z$ is the set of parameters and $n$ is the number of parameters. That is:

\begin{equation}
\chi^2_{\gamma} = F^{-1} (1-\gamma;n)
\end{equation}

When the buffer file has enough statistics, the data are used to calculate the variances and the covariances, as well as, the confidence limits for the depth profile considering the uncertainties of all parameters. The size of the buffer is user defined (3000 is set by default, but other values are also available, like the value of 40000 as recommended by the Guide of Uncertainties and Measurements in its appendix on Monte-Carlo calculation of uncertainties).

\subsection{Approach for layer selection based on Ockham's razor}

Interpretation of a set of experimental measurements requires to choose a model (ie. the depth discretization of the sample) which will fit the data to some extent out of a family of possible alternative models (like different concentration profiles or layer settings). The tempting choice of selecting the model providing the best fit (lowest $\chi^2$) is likely to over-fit the experimental data \cite{mackay}.

In MultiSIMNRA the number of layers, and consequently the number of fitting parameters, is defined by the user. Layers are not automatically created, albeit there are tools to slice, scale, merge and insert layers, making the layer management easier. However, it is known that an exaggerated number of fitting parameters leads to a reduction of the objective function, but it also introduces noise (artifacts) in the final solution by an over-fit of the experimental data. 

A principled guideline - automatically incorporating Ockham's razor - is given by a Bayesian model selection approach \cite{udo1}, which balances the goodness-of-fits of the models (or more precisely: the posterior probability distribution of the model parameters) with their complexities. The relevant quantity for this is the so called evidence \cite{linden_dose_toussaint} or marginal likelihood.

Unfortunately, despite recent algorithmic progress in computing this quantity (e.g. using Nested sampling \cite{linden_dose_toussaint,Pfeifenberger}) it is still time consuming and cumbersome to compute since the plain Markov Chain Monte Carlo (MCMC) is too inefficient. For this reason, several easier accessible approximations have been suggested. In the limit of many data and the (in that case reasonable) assumption of a multivariate Gaussian posterior distribution, the Akaike information criterion (AIC) can be derived. It depends - besides $\chi^2$ - only on the number of parameters $n$ and the number of data-points $N$ and is given by \cite{akaike}:

\begin{equation}
\label{AIC}
AIC = \chi^2 + 2n + \frac{ 2n(n+1) }{ N - n - 1 }
\end{equation}

The $AIC$ estimator is a asymptotically unbiased quantitative measure of Occam's razor. The minimum value of the $AIC$ coefficient represents the simplest description of the sample with no loss of information. Note that eq.~\ref{AIC} is proportional to the number of fitting parameters thus a reduction of $\chi^2$ is not necessarily followed by a reduction of the $AIC$ coefficient. Hence, a reduction of the $\chi^2$ function not followed by a reduction of the $AIC$ estimator, represents an over-fitting of the experimental data.

Comparing the $AIC$ estimator calculated for different description of the sample enable the quantitative comparison of different fitted models (differences regarding the number of fitting parameters) in order to judge the level of empirical support of each model (see table \ref{table_aic}).

\begin{table}[h!]
  \centering
  \caption{Differences of AIC estimators and respective level of empirical support for differentiation.}
  \label{table_aic}
  \begin{tabular}{c|c}
    \hline
    $AIC_i - AIC_{min}$ & Level of empirical support \\
    \hline
    0 - 2 & Substantial\\
    4 - 7 & Considerably less\\
    $>$ 10 & Essentially none\\
    \hline
  \end{tabular}
\end{table}

\subsection{Optimization algorithms}

At the time of this publication four algorithms for minimizing the objective function were implemented and are available as an option for the user. The first option is the Nelder-Mead’s Simplex (NMS) \cite{NELDERMEAD_simplex, CACECI_simplex}, a direct search method, computationally simple which does not require calculation of derivatives. This is the same algorithm implemented in the fitting tools of SIMNRA. It is known that NMS may become less efficient for large number of fitting parameters and that it eventually stagnates in local minima. To correct for these issues, in our implementation,  adaptive Simplex-parameters was implemented \cite{GAO_adaptivesimplex} and also oriented restarts \cite{KELLEY_orientedrestarts}, which are two features that aim to improve the effectiveness of the NMS algorithm. The second algorithm is a modified version of the Levenberg-Marquardt algorithm (LMA), which is a robust and fast deterministic optimization algorithm \cite{MARQUART}, but very sensitive to the initial parameters and does not guarantee global minimization. The third method is the evolutionary annealing-simplex algorithm \cite{EFSTRATIADIS_eas}, which is a probabilistic heuristic global optimization technique that joins ideas from different methodological approaches, where a generalized downhill simplex methodology is coupled with a simulated annealing procedure. The fourth method is a modified version of the evolutionary algorithm, Differential Evolution (DE), called Adaptive Differential Evolution with crossover rate repair (RCR-JADE) \cite{GONG_differentialevolition}. An interesting characteristic of DE is its robustness to the initial guess, but it is sensitive to the search-space boundaries. 

An initial guess for the values of concentrations of elements in the sample are informed by the user to the MultiSIMNRA software in the form of a table, together with its range limits.  The program also includes some tools to handle the sample description (like scale layer, slice layer, remove layer, etc.) that make the user interaction easier.

\begin{figure*}[!htb]
\centering
\includegraphics[width=13cm]{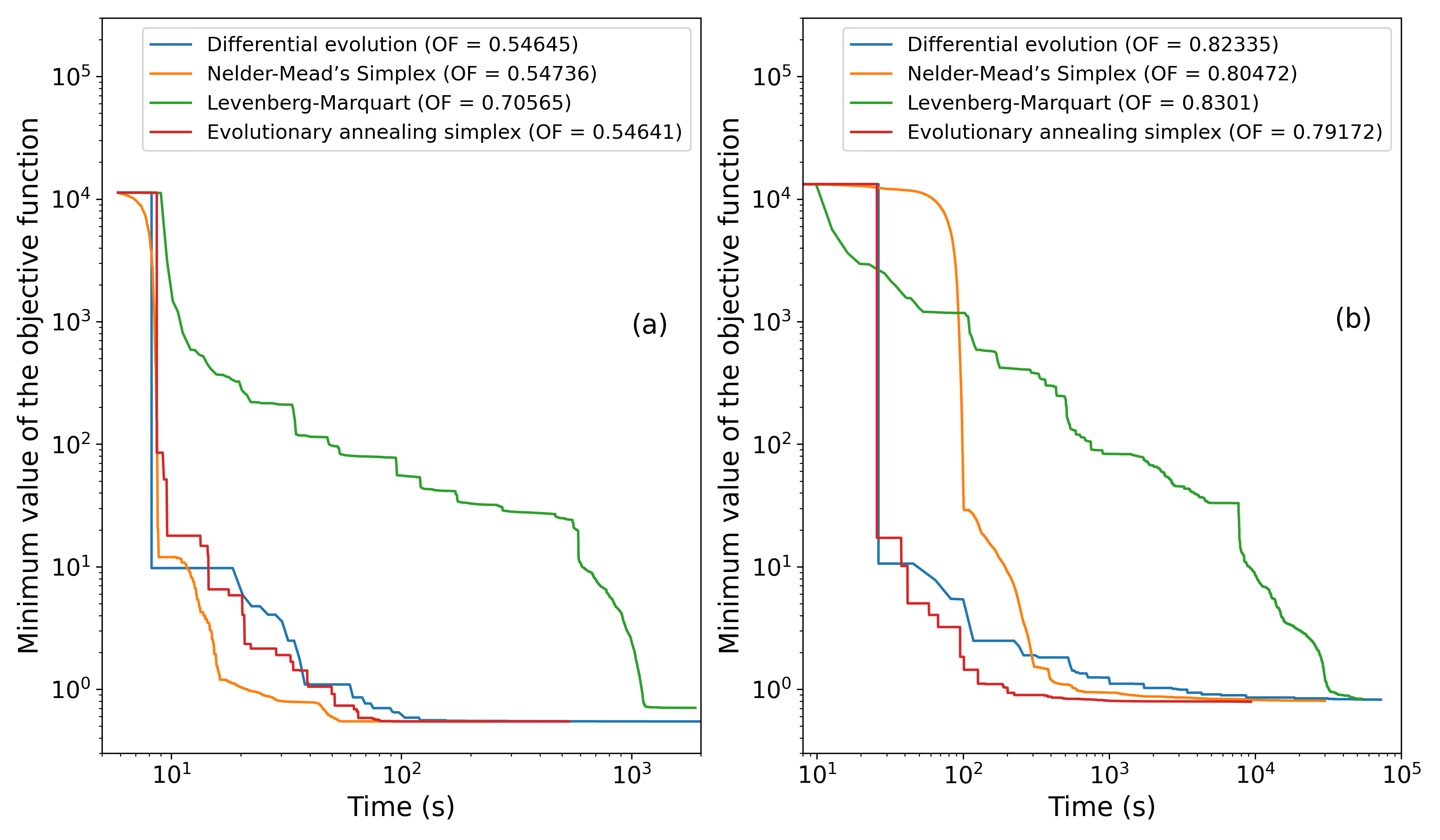}
\caption{Convergence performance of the implemented optimization algorithms. (a) for seven fitting parameters and (b) for 32 fitting parameters. The least values achieved for the objective function (OF) are shown in the legend.}
\label{fig_benchmark}
\end{figure*}

A performance test was done to compare the convergence time of all four optimization algorithms. The test consisted in using the MultiSIMNRA code to fit a set of four simulated spectra of an idealized sample (with Poisson noise added) in different configurations of energy and scattering angle. The sample consisted of a three layered model with a total of seven fitting parameters. The initial guess to start the fitting procedure was the same for all algorithms and chosen to be very far from the known solution and no human intervention was made during the optimization process. All the three algorithms converged to the known solution (with the exception of the Levenberg–Marquardt algorithm that slowly converged to a configuration just close to the known solution, probably due to the derivative instabilities generated by the inserted noise). The time evolution of the objective function for each optimization algorithm is presented in fig.~\ref{fig_benchmark}(a).

It is worth to mention that, considering the artificial noise added to the simulated spectra, a good solution was obtained for three of the four implemented algorithms in a convergence time of less than 2 minutes. Starting with a good initial guess this time can be drastically reduced.

Another performance test was executed to compare the convergence time of the four optimization algorithms in more complex analysis. This time, the model included 32 free parameters (26 for the target description and 6 for the experimental setup). Fig.~\ref{fig_benchmark}(b) shows that all the four algorithms converged to the minimum, but Leveberg-Marquart persists as the less effective algorithm. Nealder-Mead's simplex initiated the convergence with a slight delay compared to differential evolution and the evolutionary annealing simplex. Quickly converged so that the three algorithms converged in approximately the same total convergence time. 

All performance tests were done on a desktop with the last generation of x86-64 processor.

\subsection{The use of prior information}
\label{pior}

The basic motivation behind the self-consistent approach is to integrate the maximum of information obtained from different ion beam techniques. In this sense, the use of prior information in the analysis is very important and can considerably increase the quality of the solution. Butler \cite{BUTLER_criteria} proposed a three steps approach in the analysis to reduce ambiguities: i - the use of multiple detectors; ii - the information of chemical formulas; iii - the information on diffusion profiles (what Butler called thermodynamics information). We can summarize these three steps into two more generic steps: i - the combination of multiple measurements (or the Alkemade statement \cite{ALKEMADE_ambiguity}); ii - the use of prior information. The first is a generalization of the first Butler's rule, and the latter is a generalization of the second and the third Butler's rules combined: any information on the chemical state of the elements in the sample, and on the diffusion profile can be considered prior information, as is the information on the detector electronics and experimental configuration.

Prior information is used in MultiSIMNRA by the definition of constraints through the definition of Links. A Link is a feature of MultiSIMNRA that defines the relationship between the fitting parameters. The definition of Links is important not just because of the reduction of the convergence time by the direct reduction of fitting parameters, but also to establish boundary conditions and the reduction of degrees of freedom for the fitting process, taking advantage of a prior knowledge on certain parameters. The definition of Links tying the fitting parameters increases the accuracy of the analysis and is a way to deal with problems releated to the curse of dimensionality \cite{BellmanBook}.

\subsection{Basic physical modeling}

The accuracy of the simulation codes and of fundamental physics models are crucial not just for the correctness of the obtained results but also for the algorithm convergence in the objective function optimization. Deviations in the modeling of physical phenomena become more evident when combining the information contained in different spectra due to inconsistencies in the simulation of the different conditions \cite{jeynes2020_accuracy}. All this may lead to biased result or misinterpretation of data. Thus, the accuracy of the codes must be continuously confronted with experimental data. In a software inter-comparison promoted by IAEA \cite{BARRADAS_intercomparison, BARRADAS_summary}, it was demonstrated that the main differences on the simulation results of the most used codes lies on differences in the databases of fundamental physics data, such as stopping powers, cross-sections, straggling and multiple or plural scattering. Consequently, the quality of fundamental physical data are crucial for a self-consistent analysis of multiple measurements.

According to reference \cite{HPaul}, the semi-empirical approach adopted in SRIM is still the best approximation of the available experimental data for elements with $1 \leq Z \leq 92$, even though, according to the SRIM changelog, no major updates on its database were made since the 2003 version. Merit should be given to some theoretical developments \cite{CASP} but regarding the level of agreement to experimental data aiming at accurate IBA calculations and computational performance, the semi-empirical approach of SRIM is still unbeaten.  Thus,  new accurate and traceable stopping power measurements and continuous updates of stopping power databases are highly necessary.

The situation for the cross-sections database as provided by SigmaCalc \cite{GURBICH_evaluatedcrosssection} is good concerning its continuous update, and the efforts to increase the coverage of elements and the intensive tests by means of benchmarking experiments \cite{BENCH-kokkoris}. However, concerning the uncertainties and traceability, the database is still scarse \cite{SIGMACALC-uncertainty}, and restricted to a single element (Carbon). Besides, no uncertainty analysis is made on the energy reference for such database.

MultiSIMNRA has the option to use SRIM \cite{SRIM-2003} stopping powers and SigmaCalc \cite{GURBICH_evaluatedcrosssection} evaluated cross-sections for all spectra simulations, ensuring consistency in all physical models. Despite the good internal database for stopping powers of SIMNRA, the use of SRIM data is favored for traceability purposes, since its results are intensively confronted to experimental data and vice-versa \cite{SRIM-website}. For the cross-sections, the evaluated data provided by SigmaCalc can be obtained in \cite{SIGMACALC-IAEA, SIGMACALC-gurbich} and incorporated in the SIMNRA reaction list. Starting with the new SIMNRA version \cite{MAYER_simnra7}, a built-in version of SigmaCalc is implemented and the download from an external source is no longer necessary.

The use of experimental cross sections in situations where there are no evaluated data may rise the problem of how to combine different accuracies in the fundamental databases to calculate the statistical weights in the objective function (cross-sections of different elements or scattering angles, for instance). This is a topic not yet adequately addressed in the literature and therefore not considered in MultiSIMNRA.

\section{Results}

To illustrate the potential of a self-consistent approach analyzing nuclear techniques, we present two demonstration exercises. The first one is based on the processing of an artificially generated dataset, and the second discusses the analysis and the processing of several different experimental spectra.

\subsection{Butler's exercise}

\begin{figure*}[!htb]
\centering
\includegraphics[width=10cm]{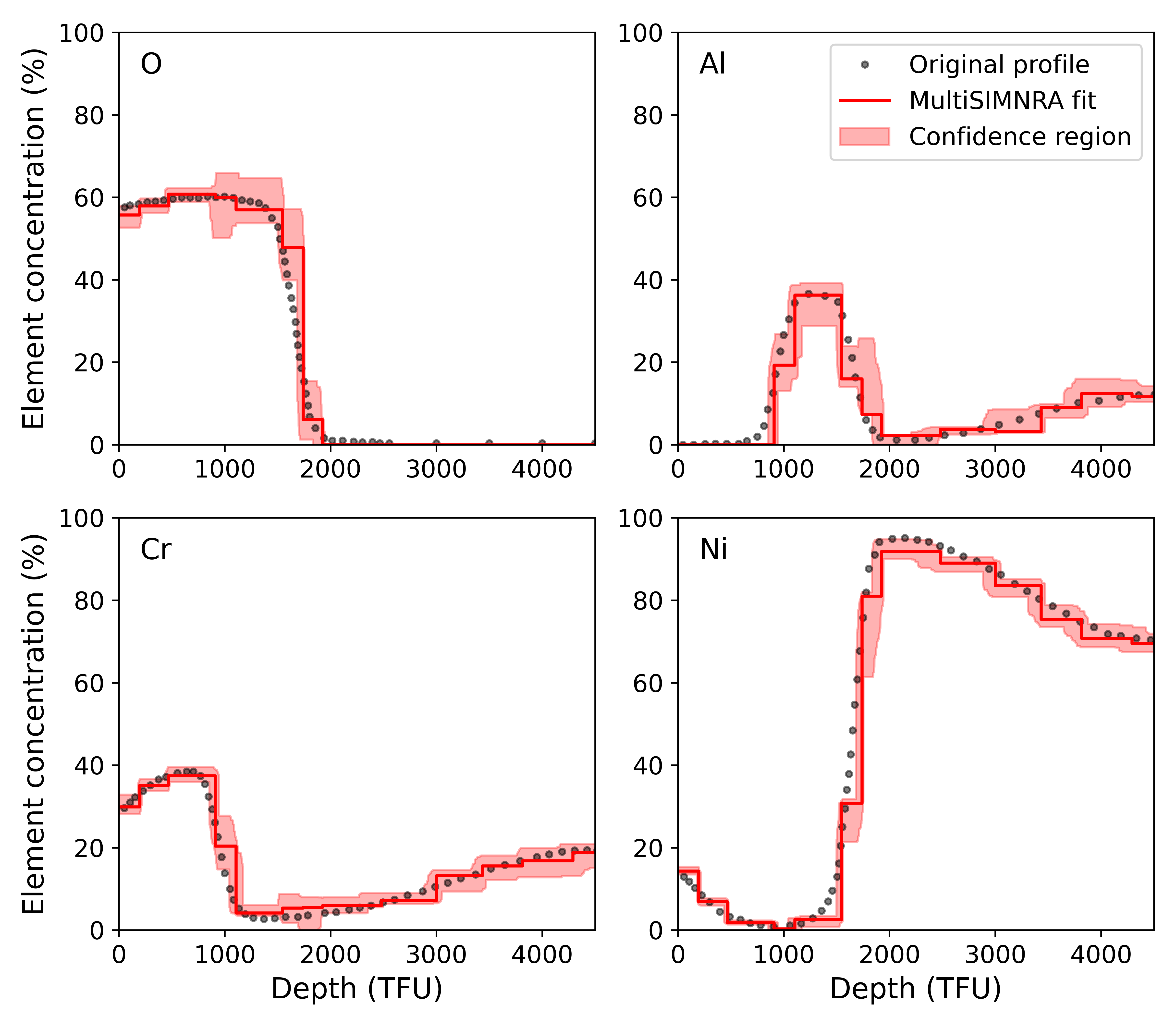}
\caption{Comparison of Butler's hypothetical sample description (dotted line) and fitted data using MultiSIMNRA (continuous line).}
\label{fig_butler_profile}
\end{figure*}

To illustrate how the ambiguities inherent to an RBS analysis can be solved, Butler \cite{BUTLER_criteria} proposed an exercise based on the fit of simulated data for an hypothetical sample of metallurgical interest. The sample contains a depth profile of the elements O, Al, Ni, and Cr (see fig. \ref{fig_butler_profile}). What is important about this exercise is that the RBS simulation provides a spectrum with overlapping signals from different elements, that opens space for ambiguities in the interpretation of the data. In this sense, Butler also proposed three steps in the analysis to avoid the ambiguities and to extend the applicability of ion beam techniques (see section on the use of prior information). This test was also successfully accomplished by Barradas using the NDF software \cite{BARRADAS_datafurnance} following the three Butler's steps. Here, we adopted a different approach: For a more realistic simulation, a Poisson noise was added to the data and, instead of using the chemical state information, we followed the Alkemade \cite{ALKEMADE_ambiguity} statement (at least $N-1$ spectra may be necessary to solve $N$ variables). Thus, without any previous knowledge of the sample, all information was extracted only by the self-consistent analysis.

\begin{figure*}[!htb]
\centering
\includegraphics[width=10cm]{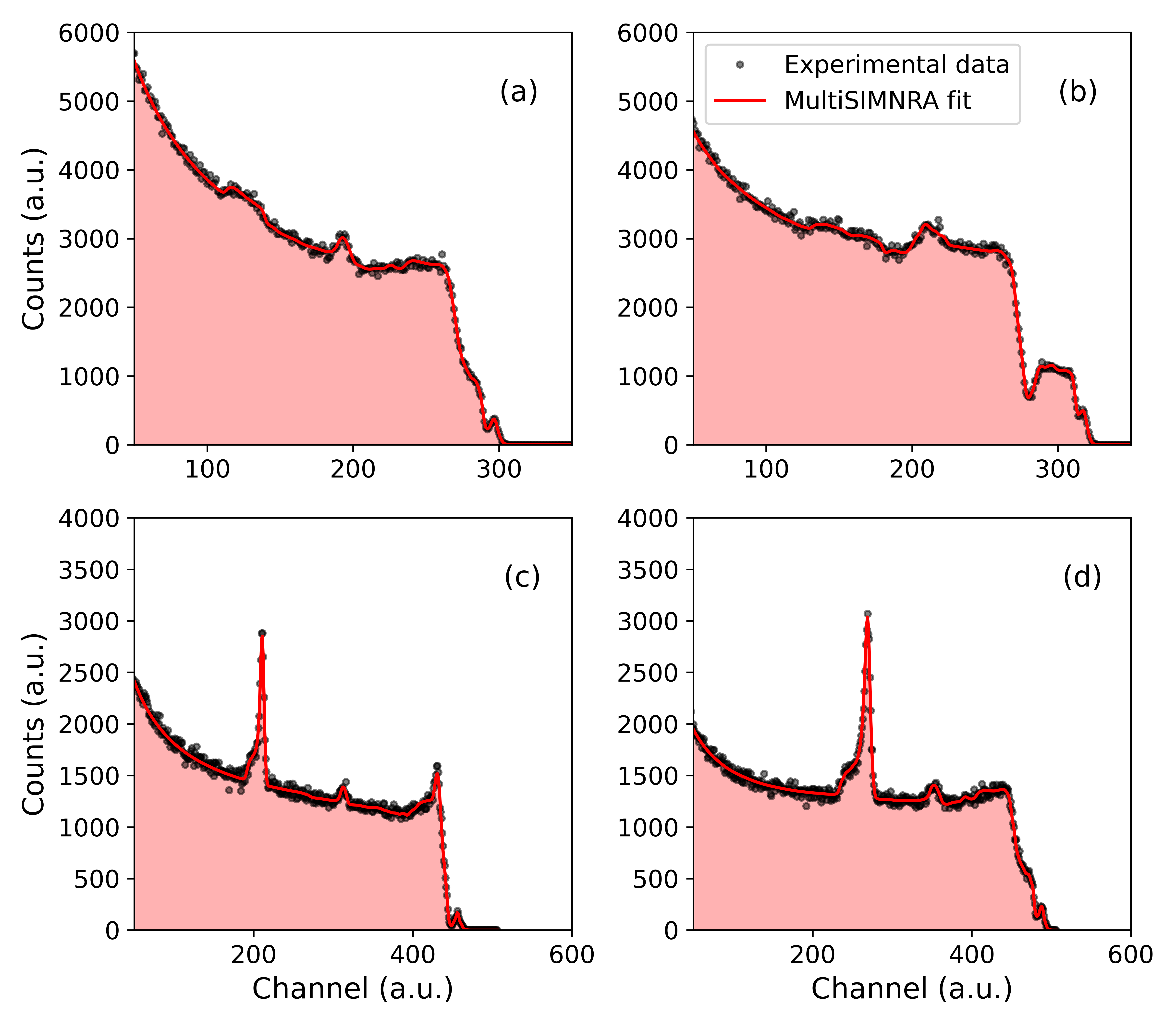}
\caption{Spectra of the Butler's hypothetical sample as an exercise of solution for ambiguities. Dots are simulated spectra with addition of Poisson noise. See text for simulation conditions. Continuous lines are the fitted curves obtained by MultiSIMNRA.}
\label{fig_butler_spectra}
\end{figure*}

Fig. ~\ref{fig_butler_spectra} shows the simulated data (continuous line) of Butler's hypothetical sample at four different experimental conditions with the addition of Poisson noise, and the superposition of the fitted curves obtained by MultiSIMNRA (discontinuous lines). The spectrum in fig. ~\ref{fig_butler_spectra}-(a) was generated at the same conditions as in the Butler's original paper: incident helium ion energy, $2000$ keV; detector angle, $170^{\circ}$; calibration, $5$ keV/ch; intercept, $25$ keV, and resolution, $15$ keV. It was calculated for a detector with $1$ msr solid angle and $10$ $\mu$C integrated charge. The spectrum in fig. ~\ref{fig_butler_spectra}-(b) was generated at the same conditions as fig. ~\ref{fig_butler_spectra}-(a) but with the detector placed at $120^{\circ}$ scattering angle. Data in fig. ~\ref{fig_butler_spectra}-(c) and (d) were generated in the same conditions as fig. ~\ref{fig_butler_spectra}-(a) and (b), respectively, but with $3050$ keV incident helium ion energy to take advantage of the resonant non-Rutherford scattering cross-section for the $^{16}$O($\alpha$,$\alpha$)$^{16}$O reaction. The cross-section used for the $^{16}$O($\alpha$,$\alpha$)$^{16}$O reaction was calculated using SigmaCalc ~\cite{GURBICH_evaluatedcrosssection}.

The depth profiles obtained by MultiSIMNRA are presented in fig.~\ref{fig_butler_profile}. The dotted lines are the data used to generate all spectra presented in fig.~\ref{fig_butler_spectra}. The continuous lines are the profiles obtained using MultiSIMNRA and the dashed painted region gives the confidence limits for the result. It is possible to observe the good agreement between the original profile and the fitted model.

One aspect worth mentioning is that uncertainty bands represent all possible profiles that passed the inference test for uncertainty estimation (see the section on uncertainty calculation). Thus the profiles and uncertainty bands are highly dependent on the number of layers of the model (see the section on model selection based on Ockham's razor). Therefore, it is not easy to represent the variances and covariances between all fitted parameters graphically. However, MultiSIMNRA provides the correlation matrix and total amounts of the elements in the final report. The latter presents minor uncertainties compared to the uncertainty bands because it considers the correlations properly.

\subsection{Fitting experimental data}

To demonstrate the advantages of joint processing of spectra obtained by multiple ion beam techniques, we present the analysis of a thin-film iron samples using heavy-ion time-of-flight RBS data and time-of-flight ERD data. The sample is a thin iron film with 0.7at.\% of tungsten, used as a model system for the EUROFER low activation alloy intended to be used in future fusion power plants. The film was sputter deposited on silicon to investigate the preferential sputtering of iron and the tungsten accumulation at the surface during low energy deuterium irradiation \cite{app_8}. Thus, top layers' depth profile characterization in a nanometer range is of high importance to this study, either for heavier (Fe and W) or for lighter (C and O) elements.

\begin{figure*}[!htb]
\centering
\includegraphics[width=11cm]{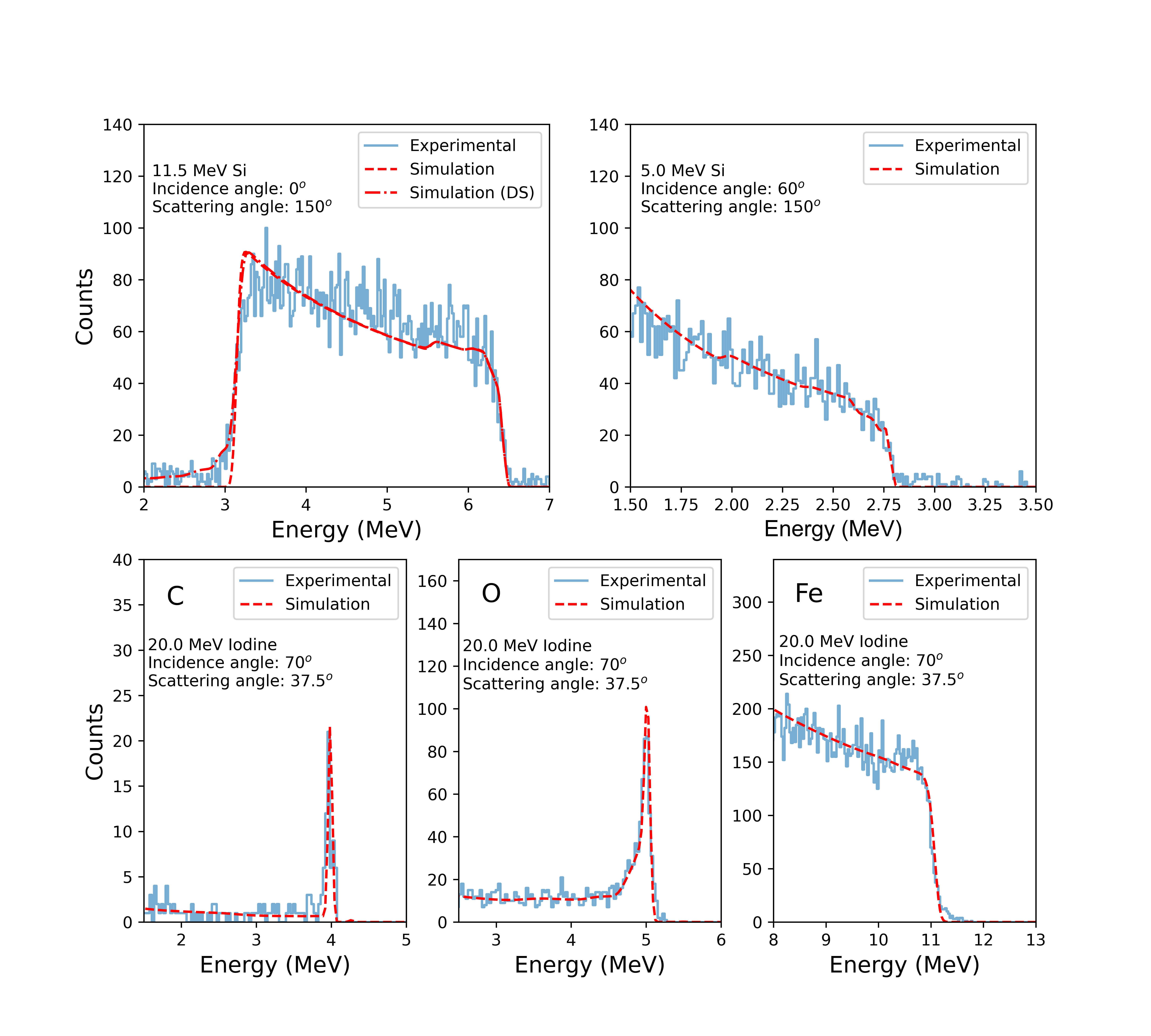}
\caption{MultiSIMNRA fit of different IBA measurements. Simulation stands for the SIMNRA calculation of the adjusted model and Simulation (DS) is the same with dual scattering algorithm enabled. }
\label{fig_epx_fit}
\end{figure*}

The tungsten accumulation at the surface of the iron films is of high interest in the design and construction of fusion reactors. For our purpose, it presents a stringent test of multiple spectra analysis by MultiSIMNRA, using two different scattering techniques in two different laboratories: the ToF-RBS data was obtained at the Max-Planck Institut für Plasmaphysik in Garching, Germany \cite{app_8}, and the ToF-ERD data was obtained at the Ruder Boscovic Institute in Zagreb, Croatia \cite{toferda_zagreb}. Fig. \ref{fig_epx_fit} shows experimental data and the simulated spectra for the sample's adjusted model.

The top plots in fig.~\ref{fig_epx_fit} are the ToF-RBS data: energy spectra of scattered 11.5 MeV silicon ions with incidence perpendicular to the surface on the left, and scattered 5.0 MeV silicon ions with an incidence angle of 60$^\circ$ on the right. In both cases, the detector was placed at 150$^\circ$ scattering angle \cite{app_8}. The bottom plots show the ToF-ERD data for each recoiled ion species: C, O, and Fe, respectively. The incident beam was 20.0 MeV iodine ions at an incidence angle of 70$^\circ$. The detector was placed at 37.5$^\circ$ scattering angle. The scattering angle is always in respect to the incident beam \cite{toferda_zagreb}.

\begin{figure}[!htb]
\centering
\includegraphics[width=7cm]{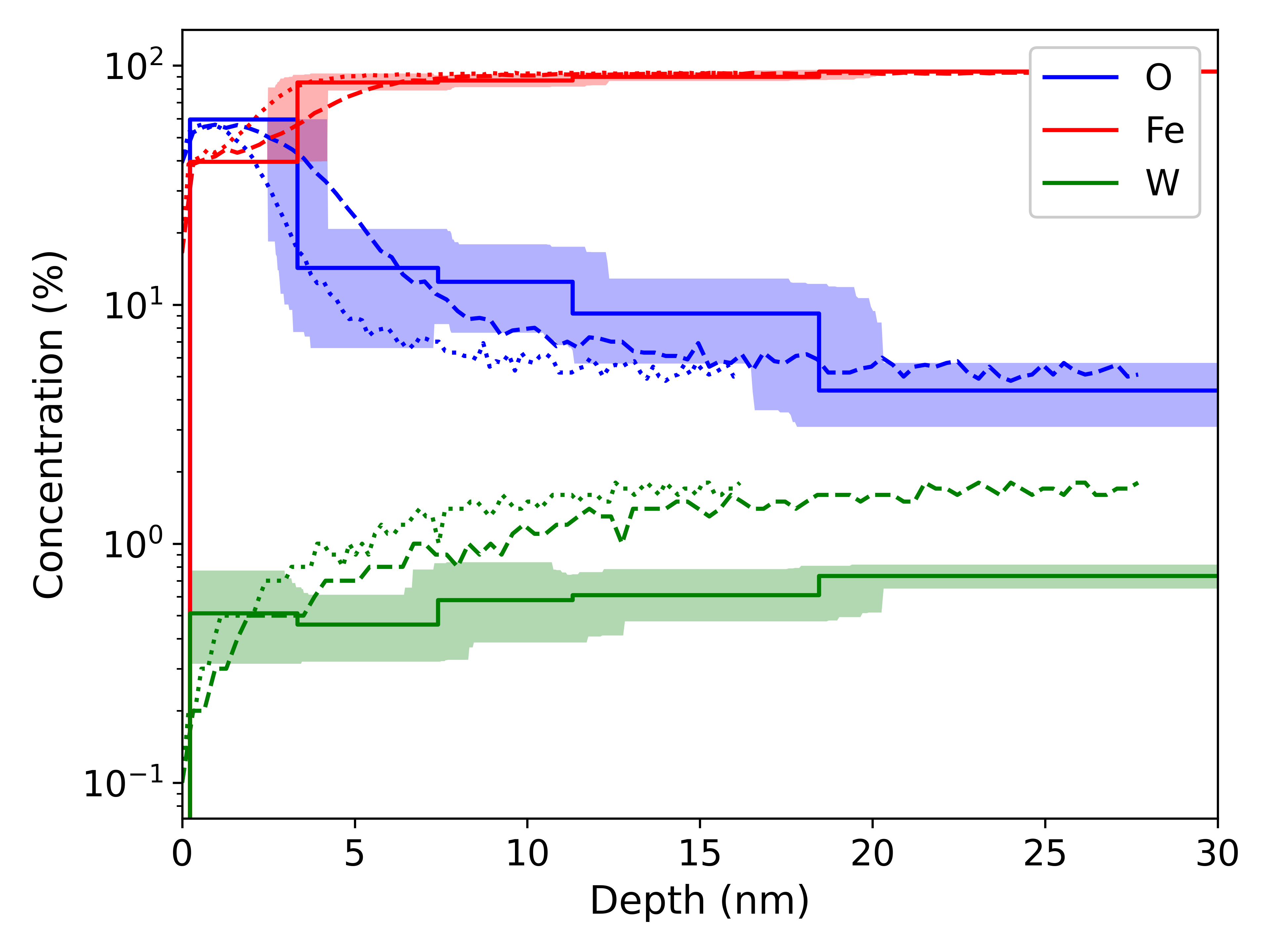}
\caption{Comparison of sputter-XPS depth profiles (dashed and dotted lines) with the results obtained using MultiSIMNRA (continuous lines) including uncertainty bands. The dashed line shows XPS depth profile with the depth scale calculated using SRIM \cite{SRIM-2003} predicted sputtering rate (with displacement energy of 25 eV for iron and tungsten, as SRIM default values), while the dotted line shows the XPS depth profiles calculated using SDTrim.SP \cite{sdtrimsp} predicted sputtering rate (with displacement energy of 25 eV for iron and 90 eV for tungsten). }
\label{fig_exp_asdeposited}
\end{figure}

The only assumptions used are the full oxidation of the iron and of the tungsten at the surface. These assumptions were included in the analysis as prior information, and imposed as a constraint by setting a Link (see section on the use of prior information) between the elements, considering Fe$_2$O$_3$ and WO chemical states at the surface. The final result of the depth profile model for elemental concentrations is finally presented in fig.~\ref{fig_exp_asdeposited} and compared with sputter-XPS data (the two discontinuous lines, differing only on the sputtering rate to correct the depth scale).

The excellent agreement between the self-consistent analysis of the ion beam methods and the sputter-XPS measurement is evident. The relative higher tungsten concentration in the XPS analysis is probably a sputtering enhancement caused by preferential sputtering of iron due to the argon primary beam (10 keV energy and incidence angle of 20$^\circ$), hence an XPS measurement artifact. This points to another positive feature of the ion beam methods: direct assessment of depth profile information with minimum sample modification. For the sputter-XPS data, the depth scale was obtained using sputtering rates as calculated by SDTrim.SP \cite{sdtrimsp} (version 6.05) SRIM \cite{SRIM-2003} (version 2013).

\section{Discussion}

The exercise using artificially generated data indicates that the self-consistent approach effectively solves ambiguities inherent to nuclear scattering analytical techniques. Including additional and independent measurements constrained the convergence of the model to the correct depth profile without any previous knowledge about the chemical states, as suggested in another study \cite{BARRADAS_datafurnance}. Additional measurements and prior information are helpful to extend the applicability of the ion beam analytical methods and may operate interchangeably. They also improve the accuracy of the material characterization provided by these techniques.

The obtained depth profiles agree very well with the profile used to generate the artificial dataset, showing no bias is introduced by the weights adopted in the objective function or the difference in counting statistics of the different measurements. Additionally, comparing both depth profiles, the uncertainty evaluation seems to be robust and well estimated. Providing an accessible and reliable method to evaluate fitting uncertainties is a positive aspect of the MultiSIMNRA software. It can be adopted to improve any analysis protocol using ion beam methods.

The exercise using experimental data, nicely shows that the self-consistent approach was very effective in finding a sample model representing the consensus among the different data, even though they were obtained using different ion beams, different experimental approaches, and even different laboratories. All this is yet more impressive, considering the uncertainties associated with the stopping power databases of heavy ions, different experimental conditions and consequently different systematic errors, including for example possible differences in accelerator calibrations or setup characterizations.

The quality of the result is confirmed by the excellent agreement between the obtained depth profile and the sputter-XPS measurements. The results fall within the confidence limits defined by the uncertainty bars, including systematic uncertainties related to the conversion of the depth scale using the SRIM calculated sputter ratios. Differences in the calculated tungsten depth profile were attributed to the preferential sputtering of other elements. The reduced damage to the sample caused by the ion beam is an additional positive aspect of the scattering techniques, since it avoids analytical artifacts in the measured depth profiles.

\section{Conclusions}

Ion beam analytical techniques are receiving a renewed interest and increased applications, primarily because of advanced analysis algorithms and the synergistic combination of multiple measurements. This approach was initiated by NDF in 2000\cite{BARRADAS_simultaneousandconsistent} and continuously improving since then. This seems to be the road to maintain nuclear techniques competitive. 

Additionally, other thin-film characterization methods hardly present the same potential of being a primary standard method featured by nuclear techniques. All these make nuclear techniques essential and, to a certain extend, irreplaceable for thin-film characterization.

Including prior information as constraints in the optimization algorithms, opens the possibility to extend the synergy with other techniques, especially those that provide information about the chemical states.

The statistical method here proposed to evaluate fitting uncertainties, proved to be robust and essential for quantitative comparison with other techniques.

The weighted-sum method for multi-objective optimization enabled the self-consistent approach. The normalization term-by-term is necessary to adjust the statistical weights and provides a convenient way to set preferences by modifying the term $a_s$ in eq. \ref{MS_OF}.

Finally, the only two software packages presently available enabling the self-consistent nuclear data processing are NDF and MultiSIMNRA. The latter uses SIMNRA as a calculation engine and inherits the quality of its simulations and user interface. Since SIMNRA is widely adopted and tested by the ion beam analysis community, MultiSIMNRA is a candidate to be incorporated in the analysis protocol in laboratories worldwide.

\section{acknowledgement}

The authors thank the financial support given by FAPESP (project number 2012/00202-0) and CNPq-INCT-FNA (project number 464898/2014-5). To V.R. Vanin to discussions on the uncertainty evaluation and the bias inserted by counting statistics. Thanks also to I. Bogdanovi\'c Radovi\'c and Z. Siketi\'c from Institut Ruder Bo\^skovi\'c in Croatia for the ToF-ERDA measurements.

\section{acknowledgement}



\bibliography{references}

\end{document}